\documentclass{iopconfser}

\usepackage{geometry}
\RequirePackage[T1]{fontenc}
\usepackage[utf8]{inputenc}
\usepackage[english]{babel}
\usepackage{graphicx}
\usepackage{bm}
\usepackage{url}
\usepackage{physics}
\usepackage{amssymb}
\usepackage{float}
\usepackage{booktabs}
\usepackage{caption}
\usepackage{subcaption}
\newcommand{\ham}[0]{\mathcal{H}}

\begin{document}

\title{Mixmaster chaos in a quantum scenario: \\ a Deformed Algebra approach}

\author{Eleonora Giovannetti$^{1}$}

\affil{$^1$Department of Physics, La Sapienza University of Rome, P.le Aldo Moro, 00185 Roma, Italy}

\email{$^1$eleonora.giovannetti@uniroma1.it}

\begin{abstract}
In this work, we address the question about the fate of chaos in the Mixmaster model when we promote the system at a quantum level. We consider Deformed Commutation Relations for the Misner anisotropic variables, whose Deformed Algebras are related to two different Quantum Gravity approaches, i.e. Loop Quantum Gravity and String Theory. Also, this approach naturally implements a form of Non-Commutativity between the space variables, i.e. the anisotropies, that live in a two-dimensional space. In particular, we consider the deformation in the semiclassical limit, where the Deformed Commutators become Deformed Poisson Brackets. Then, we derive the modified Belinskii-Khalatnikov-Lifshitz map in both cases, whose properties determine the chaotic behavior for the dynamics at a classical level. The result is that chaos is removed in both cases. In fact, depending on the sign of the deformation, the dynamics will either settle into oscillations between two almost-constant angles, or stop reflecting after a finite number of iterations and reach the singularity as one last simple Kasner solution.
\end{abstract}

 We will use natural units $\hslash=c=8\pi G=1$.
\section{Deformed Commutation~Relations}
\label{Algebras}
In this section, we present the Deformed Commutation Relations (DCRs) that will be later used to deform the anisotropy variables $\beta_\pm$ of the Bianchi IX model. They will form a deformed two-dimensional algebra 
\begin{subequations}
\label{PB}
\begin{equation}
\pb{p_+}{p_-}=0\,,
\end{equation}
\begin{equation}
\label{PBb}
\pb{\beta_+}{\beta_-}=\displaystyle(\beta_+\,p_--\beta_-\,p_+)\frac{f'(p_{\text{tot}})}{p_{\text{tot}}}\,,
\end{equation}
\begin{equation}
\pb{\beta_\pm}{p_\pm}=\delta_\pm f(p_{\text{tot}})\,.
\end{equation}
\end{subequations}
in which $\beta_\pm$ do not commute between each other anymore, as it is clear from \eqref{PBb}. The first form is called Brane Algebra because it reproduces the same modified Friedmann equation of Brane Cosmology (BC), and it is obtained by implementing
\begin{equation}
\label{Brane}
    f_\text{Brane}(p)=\sqrt{1+\mu^2p^2\,}\,.
\end{equation}
The second form is called Loop Algebra because it is inspired by Loop Quantum Cosmology (LQC). In this case, we consider
\begin{equation}
\label{Loop}
    f_\text{Loop}(p)=\sqrt{1-\mu^2p^2\,}\,.
\end{equation}
\section{Bianchi Models with Deformed Poisson~Algebras}
\label{DeformedBianchi}
The Bianchi IX dynamics is well-known to be chaotic near the initial singularity. In particular, its dynamics can be described as the one of a point particle in a triangular box that falls into the initial singularity while reflecting off the walls. This picture is known as Mixmaster dynamics, and our aim is to study how it is modified when we consider DCRs in the $(\beta_+,\beta_-,p_+,p_-)$ phase space. Instead of considering the full Bianchi IX potential, it is possible to consider one wall at a time and use the triangular symmetry to rotate the system, in~order to iterate the dynamics. This approximation corresponds to considering the Bianchi II model, whose Hamiltonian is
\begin{equation}
\mathcal{H}^{II}=\sqrt{p_+^2+p_-^2+3(4\pi^4)e^{4(\alpha-2\beta_+)}}\,.
\end{equation}
Our interest is understanding if chaos survives in the framework of DCRs. Therefore, we have to construct the modified reflection law that governs how the directions of the point universe before and after a reflection off the wall are mapped. In particular, when we implement the Brane Algebra $f_{\text{Brane}}(p)$ (see \eqref{Brane}) in \eqref{PB} to derive the modified Hamilton equations, it can be demonstrated that the resulting Belinskii-Khalatnikov-Lifshitz (BKL) map is
\begin{equation}
\label{BKLGUP}
\begin{cases}
\ham_i\,\sin\theta_i=\ham_f\,\sin\theta_f\,,\\
\displaystyle\ham_i+\frac{1}{2\mu}\text{arctanh}\left(\frac{\mu\,\ham_i\cos\theta_i}{\sqrt{1+\mu^2\ham_i^2\,}\,}\right)=\ham_f-\frac{1}{2\mu}\text{arctanh}\left(\frac{\mu\,\ham_f\cos\theta_f}{\sqrt{1+\mu^2\ham_f^2\,}\,}\right)\,,
\end{cases}
\end{equation}
in which the indexes ${i,f}$ parametrize the quantities before and after the reflection. In order to solve the dynamics, we need to implement numerical methods. \begin{figure}[H]  
\centering
\includegraphics[width=0.36\paperwidth]{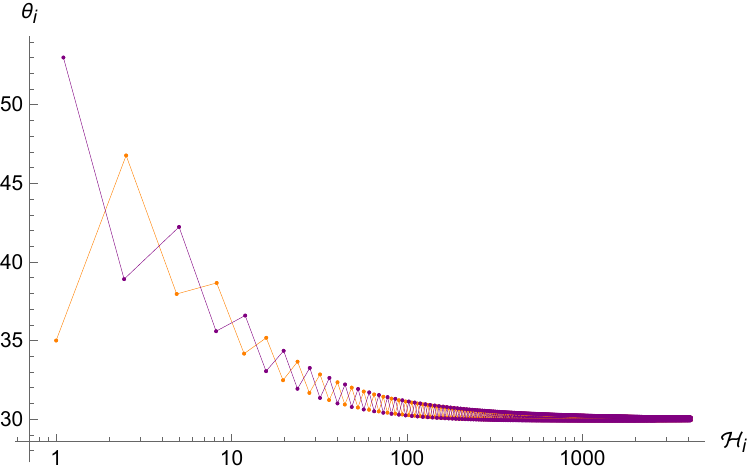}\,\,\includegraphics[width=0.36\paperwidth]{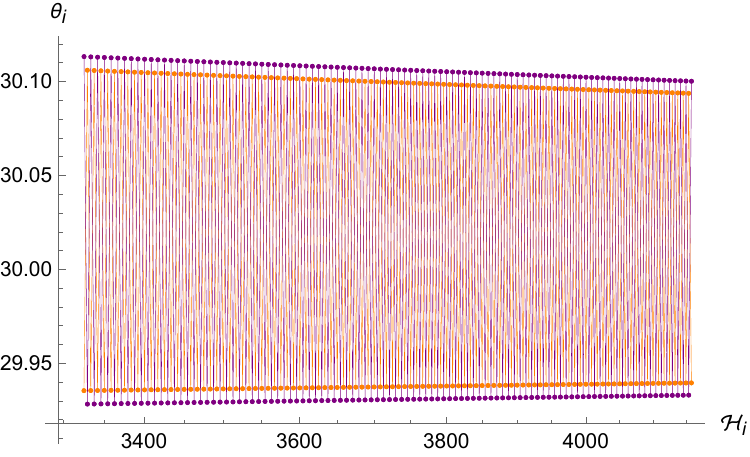}
    \caption{Left: trajectories of the point universe in the $(\mathcal{H}_i,\theta_i)$ phase space for two (sufficiently) different sets of initial conditions in the Brane Algebra. After~thousands of iterations, both converge to the same angle of $\pi/6$. Right: same figure, zoomed to the last few hundred~iterations.}
    \label{GUPb3-20000}
\end{figure} In particular, Figure~\ref{GUPb3-20000} show the trajectories in the $(\mathcal{H}_i,\theta_i)$ phase space for different initial conditions. The point particle experiences an infinite series of reflections while falling into the initial singularity. However, the~motion loses its chaotic feature and sets into an oscillatory orbit. Indeed, the~trajectories tend to become very close to each other for both very similar and sufficiently different initial conditions. In~particular, they seem to oscillate ever closer to the value $\pi/6$ which becomes an attractor value.

\begin{figure}[H]
\centering
\includegraphics[width=0.7\linewidth]{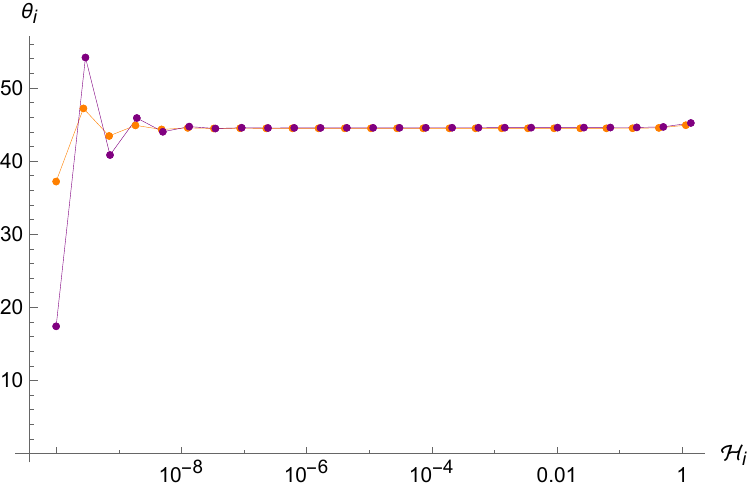}
    \caption{Trajectories 
 of the point universe in the $(\mathcal{H}_i,\theta_i)$ phase space for two different sets of initial conditions in the Loop Algebra. The~two trajectories are not sensitive to the initial conditions, and seem to converge even if they started far from each~other. Also, the~energy $\mathcal{H}_i$ keeps decreasing whereas the angle $\theta_i$ oscillates between two close values, before~stopping after 25~iterations.}
    \label{PUPe3}
\end{figure}

Now, we study the motion of the point universe in the Loop Algebra, by implementing $f_{\text{Loop}}$ (see \eqref{Loop}) in \eqref{PB}, for~which the modified BKL map is
\begin{equation}
\label{BKLLOOP}
\begin{cases}
\ham_i\,\sin\theta_i=\ham_f\,\sin\theta_f\,,\\
\displaystyle\ham_i+\frac{1}{2\mu}\arctan(\frac{\mu\,\ham_i\cos\theta_i}{\sqrt{1-\mu^2\ham_i^2\,}\,})=\ham_f-\frac{1}{2\mu}\arctan(\frac{\mu\,\ham_f\cos\theta_f}{\sqrt{1-\mu^2\ham_f^2\,}\,})\,.
\end{cases}
\end{equation}
We analyze the properties of the motion by numerically solving the iterative system, once given initial conditions $(\mathcal{H}_i,\theta_i)$. In this case, it is even more evident that the ergodicity is completely lost with respect to the Brane Algebra scenario (see Figure \ref{PUPe3}). The~system tends again to oscillate between two very close values of $\theta_i$, even with very different choices of initial conditions, before~stopping after a finite number of iterations. Furthermore, we see again that there is an attractor value, this time at $\pi/4$.

\unskip

\section{Conclusions}
We studied the Bianchi IX model in Misner variables by introducing DCRs for the anisotropies $(\beta_+, \beta_-)$, in two formulations inspired by BC and LQC. In both cases, the classical chaotic behavior disappears. In the Loop Algebra case, the discretization of $(\beta_+, \beta_-)$ implied by LQC introduces a maximum for $(p_+, p_-)$~\cite{LQC2011Review}, damping the velocity of the point universe until reflections off the potential walls cease. In the Brane Algebra case, the minimal uncertainty from BC widens the corners of the potential, enhancing the probability of escape~\cite{SebyMixmaster}. This study shows that Mixmaster chaos, while a robust classical feature, can be efficiently suppressed by quantum gravitational effects, offering semiclassical insights into the dynamics of the Bianchi IX model within BC and LQC frameworks.

\nocite{*}
\bibliographystyle{plain}
\bibliography{bibl}

\end{document}